\documentclass[twocolumn]{revtex4}

\usepackage{natbib}
\usepackage{amsmath}
\usepackage{graphicx}
\usepackage{bm}

\newcommand{\p}{\partial}

\begin{document}

\title{Strip plasmons in a two-dimensional electron gas 
with grounded electrodes}

\author{A. Satou}
\author{V. Ryzhii}
\affiliation{Computer Solid State Physics Laboratory,
University of Aizu, Aizu-Wakamatsu 965-8580, Japan}

\begin{abstract}
Plasmons in two-dimensional electron gas (2DEG) strips with grounded 
electrodes (a gate or side contacts) are investigated. 
We consider two systems: (a) the 2DEG strip with a highly conducting
gate and (b) the 2DEG strip with semi-infinite highly conducting 
side contacts. The systems are
described by hydrodynamic equations coupled with the Poisson equation.
Dispersion relations and ac electric potential 
distributions are obtained. We find the plasmon modes
whose potential distributions, and hence electron density, are localized
to the edges of the strip when the wave number of plasmons in the
direction parallel to the strip is large. Frequencies of these strip
modes are lower than those of infinite two-dimensional plasmons.
The presence of grounded electrode(s) significantly modifies the plasmon
dispersion relations.
\end{abstract}

\maketitle

\section{Introduction}

The spectrum of plasmons in electron systems
depends on their dimensionality.
In a bulk electron gas, the plasmon frequency,
$\omega$, is virtually independent of
the wave vector, $\bm{q} = (q_x,q_y,q_z)$: 
$\omega \simeq \sqrt{4\pi e^2 n/m\ae}$,
where $e$ and $m$ are the electron charge and effective mass,
respectively, $\ae$ is the dielectric constant, and $n$ is
the electron volume concentration. The spread in the electron 
velocities associated with the thermal movement
of electrons or the degeneration of the electron system results in
some dependence of $\omega$ on $\bm{q}$. However this dependence is
rather weak~\cite{LandauLifshitz-PhysicalKinetics}.
Stern~\cite{Stern-PRL-1967} 
and Chaplik~\cite{Chaplik-SPJETP-1972} (see also early papers
by Ritchie~\cite{Ritchie-PR-1957} and
Ferrell~\cite{Ferrell-PR-1958}) discussed
plasmons in the two-dimensional electron gas (2DEG) 
in the $n$-type inversion layer
of metal-insulator-semiconductor structure.
The dispersion relation for plasmons
in 2DEG is given by~\cite{Stern-PRL-1967} 
$\omega \propto \sqrt{q}$.
Here 
$\bm{q} = (q_x, q_y)$ ($z$-axis is directed perpendicular to
the 2DEG), and $q = |\bm{q}|$. 
Fetter~\cite{Fetter-PRB-1974} studied plasmons
in the classical 2DEG on the surface of liquid He.
The plasmons propagating along a one-dimensional
wire (in the $x$-direction)
correspond to~\cite{Ferrell-PRL-1964, Gold-PRB-1990} 
$\omega \propto \sqrt{\ln (2/qa)}q$,
where
$q = q_x$ and $a$  is the wire radius.
The linear dispersion relation of 2DEG with a metallic gate
has been found in Refs.~\onlinecite{Chaplik-SPJETP-1972} and 
\onlinecite{Eguiluz-PRB-1975}.
The first experimental observation of 2D plasmons in the 2DEG on
liquid He was reported
by Grimes and Adams~\cite{Grimes-PRL-1976}. Thereafter,
Allen \textit{et al.}~\cite{Allen-PRL-1977} and
Tsui \textit{et al.}~\cite{Tsui-SSC-1980} observed
far-infrared absorption and emission, respectively,
in silicon inversion layers. 

Edge plasmons in 2DEG with applied magnetic field, 
so-called edge magnetplasmons, were observed by
Mast \textit{et al.}~\cite{Mast-Physica-1984,Mast-PRL-1985} 
and Glattli \textit{et al.}~\cite{Glattli-PRL-1985}
Following their works, 
many theoretical investigations of edge magnetplasmons 
in bounded structures
(semi-infinite half plane~\cite{Wu-PRL-1985, Fetter-PRB-1985, 
Fetter-PRB-1986-Mar, Volkov-SPJETP-1988, Xia-PRB-1994}, 
disk~\cite{Fetter-PRB-1986-Apr, Volkov-SPJETP-1988},
and strip~\cite{Cataudella-PRB-1987, Volkov-SPJETP-1988})
were conducted using hydrodynamic electron transport 
models for 2DEGs
coupled with self-consistent Poisson equation.
The equations of these models produce integral equations
which can be tackled by replacing their kernels with ones,
or by expanding their solutions over complete sets of orthogonal 
functions (for comparison of these two methods see, 
for example, works by 
Fetter~\cite{Fetter-PRB-1985, Fetter-PRB-1986-Mar}).
Edge plasmons in semi-infinite half plane with gate(s) were 
discussed by Fetter~\cite{Fetter-PRB-1985, Fetter-PRB-1986-Mar}
and Nazin \textit{et al.}~\cite{Nazin-SPJETP-1987}.
The discussion of early results on 2D plasmons can be found in 
the review papers by Theis~\cite{Theis-SS-1980} and 
Ando \textit{et al.}~\cite{Ando-RMP-1982}

Rudin and Dyakonov~\cite{Rudin-PRB-1997} considered the
edge plasmons in the strip (strip plasmons)
with zero magnetic field using a variational method, and
found that two strip plasmon modes can exist in the strip structure
(in Ref.~\onlinecite{Rudin-PRB-1997} 
the lowest symmetric and antisymmetric modes were
considered). However, to use their
method, it is necessary to choose appropriate trial functions
which are usually unknown and hard to find a priori, 
especially for complex structures with grounded electrodes.

In this paper, we calculate the spectra of plasmons
propagating along strips of the 2DEG with closely located
highly conducting grounded electrodes
(a gate or side contacts), using the method
developed in Refs.~\onlinecite{Ryzhii-JAP-2004} and
\onlinecite{Satou-JJAP-2005}, in which 
the solution of the integral equation reduces to
the eigenvalue problem.
For simplicity and bearing in mind 2DEG in different systems
like field-effect transistors, 
we consider 2DEG without magnetic field
assuming the uniform electron equilibrium density
and neglecting the gradient of pressure terms in 
the equation of motion~\cite{Rudin-PRB-1997}.
We also neglect the electron collisions in 2DEG, which naturally
leads to some modification of the plasmon dispersion 
relation~\cite{Ryzhii-JAP-2004} and results in some damping.
 The specifics of the geometry of
the 2DEG systems in question and the charges
induced by the perturbations of the electron
density in the metallic region results in a modification
of the plasmon spectra. We find the spatial distributions
of the ac electric potential and the dependence of their spectra
on the geometric parameters of the 2DEG system.
In particular, we show that at sufficiently large
wave numbers of the plasmons, their ac electric
potential can be localized near both edges of the strip.
For the 2DEG strip with metallic
semi-infinite half-plane electrodes on its both sides,
we find the higher strip plasmon modes.

Since the 2DEG systems similar to those studied in the paper
can serve as resonant  cavities and wave guides in different
heterostructure devices  operating in the terahertz range 
of frequencies~\cite{Shur-HSES-2003},
the results obtained below can be helpful
of the development and optimization of such devices.

\section{Equations of the model}

\begin{figure}[hb]
 \begin{center}
  \includegraphics[height=3cm]{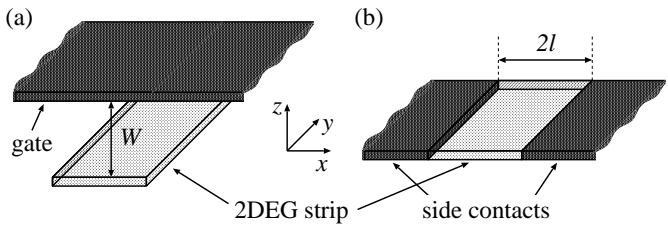}
  \caption{\label{Fig-schematic} Schematic views of structures under
  consideration: (a) gated cavity and (b) slot diode.}
 \end{center}
\end{figure}

We consider the following systems:
(a) the 2DEG strip with 
a highly conducting gate of spacing $W$ between them (gated cavity)
and (b) the 2DEG strip with 
semi-infinite highly conducting
side contacts (slot diode).
The structure considered in Refs.~\onlinecite{Volkov-SPJETP-1988},
\onlinecite{Cataudella-PRB-1987}, and
\onlinecite{Rudin-PRB-1997}
can be considered as the structure (a) with $W\to\infty$.
Schematic views of these structures are shown in
Fig. \ref{Fig-schematic}.

We use the hydrodynamic equations (the continuity equation
and the Euler equation)
to describe the electron transport:
\begin{equation}\label{Eq-ContinuityEq}
  \frac{\p\Sigma}{\p t}+\nabla\cdot(\Sigma\bm{u}) = 0,
\end{equation}

\begin{equation}\label{Eq-EularEq}
  \frac{\p\bm{u}}{\p t}+(\bm{u}\cdot\nabla)\bm{u} = 
  \frac{e}{m}\nabla\varphi,
\end{equation}
where $\Sigma = \Sigma(x, y, t)$ is
the electron sheet concentration in the strip, 
$\bm{u} = \bm{u}(x, y, t)$ is the electron 
velocity, $\varphi = \varphi(x, y, z, t)$ 
is the electric potential, and $e$ and $m$ are electron charge and
effective mass, respectively. $\bm{u}$ is in $xy$-plane and
$\nabla = (\p/\p x, \p/\p y)$.
The hydrodynamic equations are coupled with 
the Poisson equation:
\begin{equation}\label{Eq-PoissonEq}
  \frac{\p^2 \varphi}{\p x^2}
  +\frac{\p^2 \varphi}{\p y^2}
  +\frac{\p^2 \varphi}{\p z^2}
  =
  \frac{4\pi e}{\ae}\Sigma\delta(z)\theta(l+x)\theta(l-x),
\end{equation}
where $\ae$ is the dielectric constant of the medium
surrounding the strip, $\delta$ is the Dirac delta function,
and $\theta$ is the Heaviside step function.

We consider the traveling wave with frequency $\omega$
and wave number $q_y$ along $y$-axis, assuming that the potential
(as well as other quantities) comprises both the dc
and ac components: $\varphi(x,y,z,t) = \varphi_0(x,z)+
\varphi_{\omega}(x,z)\exp[i(q_yy-\omega t)]$.
Then, linearized version of Eqs.~(\ref{Eq-ContinuityEq})
and (\ref{Eq-EularEq}) together with Eq.~(\ref{Eq-PoissonEq})
gives the following self-consistent equation:
\begin{eqnarray}\label{Eq-SelfConsistentEq}
  & &
   \frac{\p^2 \varphi_{\omega}}{\p x^2}
   +\frac{\p^2 \varphi_{\omega}}{\p z^2}
   -q_y^2\varphi_{\omega}
   \nonumber \\
 & = &
  \frac{2l}{\pi\lambda_{\omega}}
  \left(
   \frac{\p^2 \varphi_{\omega}}{\p x^2}
   -q_y^2\varphi_{\omega}
  \right)
  \delta(z)\theta(l+x)\theta(l-x),
\end{eqnarray}
where  $\lambda_{\omega} = \omega^2/\Omega^2$ and
$\Omega = \sqrt{2\pi^2e^2\Sigma_0/m\ae l}$ is the characteristic
plasma frequency.
Equation~(\ref{Eq-SelfConsistentEq}) can be reduced to 
the following:
\begin{eqnarray}\label{Eq-Potential}
  \varphi_{\omega}(x, z)  & = &
   -\frac{2l}{\pi\lambda_{\omega}}
   \int_{-l}^{l}dx'G(x, z; x', 0)
   \nonumber \\
 & &
  \times \left(\frac{\p^2}{\p x'^2}-q_y^2\right)
  \varphi_{\omega}(x', 0),
\end{eqnarray}
where $G(x, z; x', z')$ is the Green function which can be determined
by the boundary conditions for each structure (boundary conditions and 
the Green function for each 
structure is introduced in subsequent sections).

For $|x|\le l$ and $z=0$, Eq.~(\ref{Eq-IntegralEq})
is reduced to the following integral equation:
\begin{equation}\label{Eq-IntegralEq}
 \varphi_{\omega}(x)  = 
   -\frac{2l}{\pi\lambda_{\omega}}
   \int_{-l}^{l}dx'G(x, 0; x', 0)
   \left(\frac{\p^2}{\p x'^2}-q_y^2\right)
  \varphi_{\omega}(x'),
\end{equation}
where $\varphi_{\omega}(x) = \varphi_{\omega}(x, 0)$.
To solve Eq.~(\ref{Eq-IntegralEq}), we
expand $\varphi_{\omega}(x)$ in the cosine and sine series 
corresponding to symmetric and antisymmetric modes of potential,
respectively. The series are different for 
each structure because boundary conditions at $|x| = l$ are
different.
Then, we can reduce the equation to the 
eigenvalue problem with an infinite matrix
whose matrix elements are determined by the Green
function $G$ and the expansion.
The problem can be solved numerically
by truncating the matrix up to size $N$, evaluating 
its elements numerically, and then solving the eigenvalue problem
with the finite matrix.
 Convergence of eigenvalues
and eigenvectors was checked by increasing $N$, and we use the value
$N = 100$ for all calculations done in the paper.
An eigenvalue of the matrix, $\lambda_{\omega}$,
gives the plasma frequency of a mode
\begin{equation}\label{Eq-DispersionRelation}
 \omega = \sqrt{\lambda_{\omega}}\Omega.
\end{equation}

\section{Strip plasmons in a gated cavity}

Since electrons in the 2DEG strip of a gated cavity
cannot go outside, the $x$-component of their velocity 
at edges of the strip
must be zero, i.e., in terms of the potential,
$\p\varphi_{\omega}/\p x|_{z=0, |x|=l} = 0$.
Consequently, the Green function for a gated cavity is
\begin{eqnarray}\label{Eq-GreenFunction-GC}
 & & G^{(GC)}(x, z; x', z') 
  \nonumber \\
 & = &
  \frac{1}{2\pi}K_0
  \left(q_y\sqrt{(x-x')^2+(z-z')^2}\right)
  \nonumber \\
 & & -\frac{1}{2\pi}K_0
  \left(q_y\sqrt{(x-x')^2+(z-z'-2W)^2}\right),
\end{eqnarray}
where the first term of Eq.~(\ref{Eq-GreenFunction-GC})
is the Green function for a free cavity.
The expansions which satisfy the boundary conditions are the
following:
\begin{equation}\label{Eq-ExpansionS-GC}
 \varphi^{(s)}_{\omega}(x) = \frac{c_0}{\sqrt{2}}+
  \sum_{k=1}^{\infty}c^{(s)}_k\cos(q_{2k}x)
\end{equation}
for symmetric modes and
\begin{equation}\label{Eq-ExpansionA-GC}
 \varphi^{(a)}_{\omega}(x) = \sum_{k=1}^{\infty}c^{(a)}_k
  \sin(q_{2k-1}x)
\end{equation}
for antisymmetric modes, where $q_{n} = n\pi/2l$. 
The effective wave number corresponding to the $x$-direction 
is equal to $q_{2n}$ for $n$-th symmetric mode (as a special case, 
zero-th mode correspond to $q_{2n} = 0$) and 
to $q_{2n-1}$ for $n$-th
antisymmetric mode.
Substituting Eqs.~(\ref{Eq-ExpansionS-GC}) and
(\ref{Eq-ExpansionA-GC}) into Eq.~(\ref{Eq-IntegralEq}) and
using the orthonormality of each term in expansions,
we arrive at the following eigenvalue problem:
\begin{equation}\label{Eq-EigenvalueProblem-GC-S}
 \sum_{k'=0}^{\infty} \theta^{(s)}_{kk'}c^{(s)}_{k'} 
  = \lambda^{(s)}_{\omega} c^{(s)}_{k}
\end{equation}
for symmetric modes, where $k=0, 1, 2, \cdots$ and
\begin{equation}\label{Eq-MatrixElement-GC-1}
 \theta^{(s)}_{00} = \frac{1}{\pi}(q_yl)^2\int_{-1}^{1}d\xi
  \int_{-1}^{1}d\xi'G^{(GC)}(\xi, \xi'),
\end{equation}
\begin{eqnarray}\label{Eq-MatrixElement-GC-2}
 \theta^{(s)}_{k0} & = &
  \frac{\sqrt{2}}{\pi}(q_yl)^2\int_{-1}^{1}d\xi
  \int_{-1}^{1}d\xi' G^{(GC)}(\xi, \xi')
  \nonumber \\
 & & \times \cos(q_{2k}l\xi)
\end{eqnarray}
when $k>0$, 
\begin{eqnarray}\label{Eq-MatrixElement-GC-3}
 \theta^{(s)}_{0k'} & = &
  \frac{\sqrt{2}}{\pi}[(q_{2k'}l)^2+(q_yl)^2]\int_{-1}^{1}d\xi
  \int_{-1}^{1}d\xi' G^{(GC)}(\xi, \xi')
  \nonumber \\
 & & \times \cos(q_{2k'}l\xi')
\end{eqnarray}
when $k'>0$, and
\begin{eqnarray}\label{Eq-MatrixElement-GC-4}
 \theta^{(s)}_{kk'}
  & = & \frac{2}{\pi}\left[(q_{2k'}l)^2+(q_yl)^2\right]
  \int_{-1}^{1}d\xi\int_{-1}^{1}d\xi'
  G^{(GC)}(\xi, \xi')
  \nonumber \\
 & & \times \cos(q_{2k}l\xi)\cos(q_{2k'}l\xi')
\end{eqnarray}
when $k, k' > 0$.
For antisymmetric modes, we obtain
\begin{equation}\label{Eq-EigenvalueProblem-GC-A}
 \sum_{k'=1}^{\infty} \theta^{(a)}_{kk'}c^{(a)}_{k'}
  = \lambda^{(a)}_{\omega} c^{(a)}_{k},
\end{equation}
where $k=1, 2, 3, \cdots$ and
\begin{eqnarray}\label{Eq-MatrixElement-GC-5}
 \theta^{(a)}_{kk'}
  & = & 
  \frac{2}{\pi}\left[(q_{2k'-1}l)^2+(q_yl)^2\right]
  \int_{-1}^{1}d\xi\int_{-1}^{1}d\xi'G^{(GC)}(\xi, \xi')
  \nonumber \\
 & & 
  \times \sin(q_{2k-1}l\xi)\sin(q_{2k'-1}l\xi').
\end{eqnarray}
Here we use the notation 
$G^{(GC)}(\xi, \xi') = G^{(GC)}(l\xi, 0; l\xi', 0)$.
The details of numerical computation of 
Eqs.~(\ref{Eq-MatrixElement-GC-1})-(\ref{Eq-MatrixElement-GC-4})
and (\ref{Eq-MatrixElement-GC-5}) are discussed in the Appendix.

\begin{figure}[ht]
 \begin{center}
  \includegraphics[height=6cm]{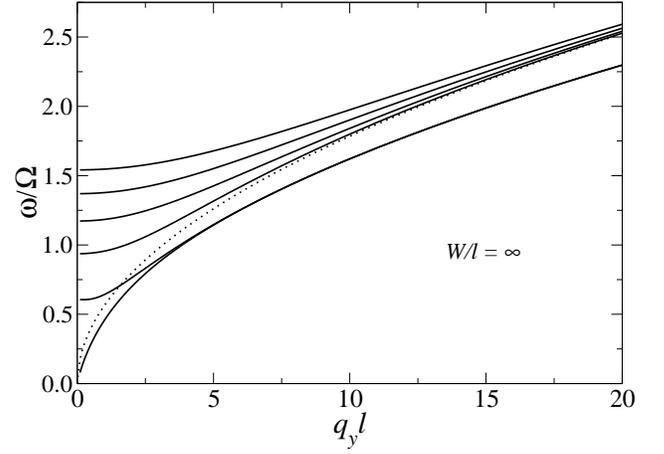}
  \caption{\label{Fig-dispersionGC-Inf}Dispersion relation for a 
  free cavity ($W/l\to\infty$) with $q_n l/\pi = 0,\, 1/2,\, 1,\, 
  3/2,\, 2,\, 5/2$. 
  Solid lines with lowest frequencies through highest ones
  correspond to $q_nl/\pi = 0$ through $q_nl/\pi = 5/2$.
  The dotted line is the dispersion relation of infinite 2D plasmons. 
  Two modes whose frequencies are lower than those of infinite 
  2D plasmons for $q_y l \gg 1$ are identified as strip
  plasmon modes.}
 \end{center}
\end{figure}

\begin{figure}[ht]
 \begin{center}
  \includegraphics[height=7cm]{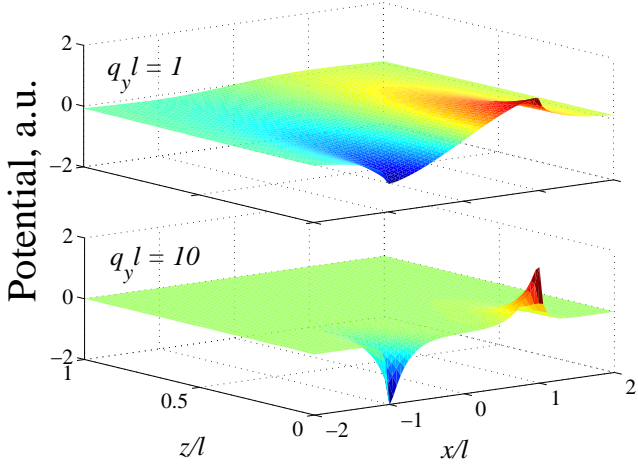}
  \caption{\label{Fig-potGC} Potential distributions of lowest 
  antisymmetric mode for a free cavity with $q_y l = 1$ (above) 
  and $q_y l = 10$ (below). The localization of potential at 
  the edges of the 2DEG strip becomes apparent for $q_y l \gg 1$.}
 \end{center}
\end{figure}

Figure \ref{Fig-dispersionGC-Inf} shows the dispersion relation
calculated for a free
cavity. The dotted line is the dispersion relation
of infinite 2D plasmons which is given by
$\omega_{p} = \sqrt{q_y l/\pi}\Omega$.
Frequencies in the limit $q_y l \to 0$ coincide with those
calculated for standing plasma waves in a free 
cavity in Ref.~\onlinecite{Satou-JJAP-2005}.
It can be seen from Fig.~\ref{Fig-dispersionGC-Inf} that 
the lowest symmetric and antisymmetric modes have lower frequencies
than those of infinite 2D plasmons for $q_y l \gg 1$, 
while frequencies of higher modes approach to those of
infinite 2D plasmons. These two modes are identified as 
strip plasmons, as is also evident from their potential 
distributions (Fig. ~\ref{Fig-potGC}).
This phenomenon was already discussed theoretically in the
case of a semi-infinite half-plane~\cite{Fetter-PRB-1986-Mar} and
in the case of a free cavity~\cite{Rudin-PRB-1997}.
For $q_y l \gg 1$, the effect of one edge on another can be
neglected, and the difference between symmetric and antisymmetric
modes becomes unimportant. The lowest modes in our case are
equivalent to an edge mode in the case of a half-plane.
The dispersion curves for lowest symmetric and antisymmetric
modes are identical with those obtained in 
Ref.~\onlinecite{Rudin-PRB-1997}. The higher modes
seem to correspond to the normal wave-like modes in $x$-direction whose
frequencies are proportional to  $\sqrt{q_n^2+q_y^2}$
because the effective wave number $q_n$ can be considered to take
continuous value when $q_y l \gg 1$.

\begin{figure}[ht]
 \begin{center}
  \includegraphics[height=5cm]{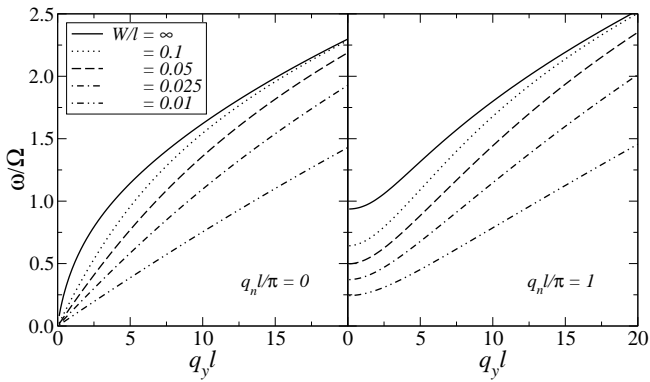}
  \caption{\label{Fig-dispersionGC-Finite} Dispersion 
  relation for a gated cavity with $q_nl/\pi = 0,\, 1$
  and $W/l = 0.01,\, 0.025,\, 0.05,\, 0.1,\, \infty$.}
 \end{center}
\end{figure}

\begin{figure}[ht]
 \begin{center}
  \includegraphics[height=6cm]{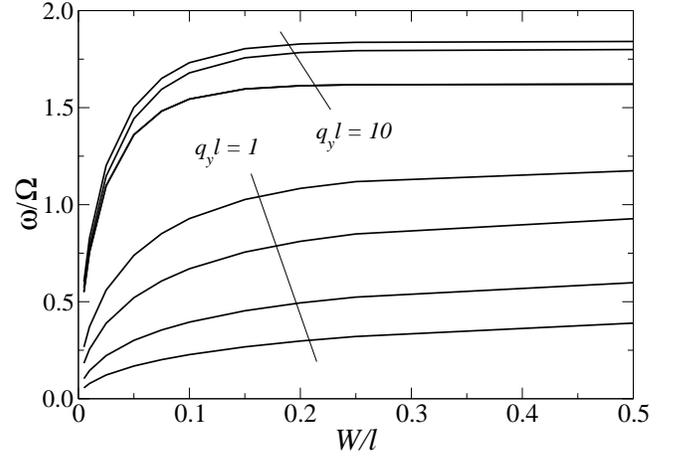}
  \caption{\label{Fig-dispersionGC-aDep} Plasma frequency
  vs the ratio $W/l$ with $q_yl = 1,\, 10$ and
  different $q_nl/\pi$ for a gated cavity.
  For each $q_yl$, four solid lines with lowest frequenciy
  through highest ones correspond to   
  $q_nl/\pi = 0,\, 1/2,\, 1,\, 3/2$. For $q_yl = 10$, 
  curves for $q_nl/\pi = 0$ and $q_nl/\pi = 1/2$ are
  indistinguishable.}
 \end{center}
\end{figure}

The dispersion relation for plasmons in a gated cavity
is illustrated in Fig.~\ref{Fig-dispersionGC-Finite}.
For the zero-th symmetric mode ($q_nl/\pi = 0$), 
one can see that, as $W/l$ decreases, the plasma frequencies decrease
and their dependence on $q_yl$ becomes linear. When $q_y \gg 1/W$, 
the effect of the gate is small and the frequencies approache
to those for a free cavity (see curves for $W/l = 0.1$).
For the first symmetric mode ($q_nl/\pi = 1$) the dependence
is more complicated due to nonzero frequency of plasmons in the limit
$q_yl \to 0$ and its dependence of $W/l$~\cite{Satou-JJAP-2005}.
Figure~\ref{Fig-dispersionGC-aDep} shows the plasma frequencies
as a function of $W/l$ for fixed $q_yl$.
For $q_yl = 1$, the influence of the gate on the frequencies
is gradual, i.e., the frequencies slowly decrease with
decreasing $W/l$ and tend to zero as $W/l \to 0$.
On the contrary, for $q_yl = 10$, the change in the frequency 
from zero to the value corresponding to $W/l\to\infty$ 
is more rapid. It is also worth 
mentioning that for $q_yl = 1$ the frequency differences between
lowest and second lowest symmetric and antisymmetric modes
are significant whereas they are not the case for $q_yl = 10$.

\section{Edge plasmons in a slot diode}

For a slot diode, the boundary conditions are given by
$\varphi_{\omega}|_{|x|\ge l, z=0} = 0$, and the relevant Green function
is
\begin{eqnarray}\label{Eq-GreenFunctionSD}
 & & 
  G^{(SD)}(x, z; x', z') 
  \nonumber \\
 & = &
  \frac{1}{2\pi}\sum_{n=-\infty}^{\infty} K_0
  \left(
   q_yl\sqrt{(\theta-\theta'-2n\pi)^2+(\psi-\psi')^2}
  \right)
  \nonumber \\
 & & - \frac{1}{2\pi}\sum_{n=-\infty}^{\infty} K_0
  \left(
   q_yl\sqrt{(\theta+\theta'-2n\pi)^2+(\psi-\psi')^2}
  \right),\nonumber \\
\end{eqnarray}
where $\theta+i\psi = \cos^{-1}[(x+iz)/l]$ and
$\theta'+i\psi' = \cos^{-1}[(x'+iz')/l]$. When evaluating
Eq.~(\ref{Eq-GreenFunctionSD}) numerically, series are cut at
appropriately large $n$, so that the $n$-th term can be
neglected. Using the expansion
\begin{equation}
 \varphi^{(s)}_{\omega}(x) = \sum_{k=1}^{\infty}
  c^{(s)}_k\cos(q_{x,2k-1}x)
\end{equation}
for symmetric modes and
\begin{equation}
 \varphi^{(a)}_{\omega}(x) = \sum_{k=1}^{\infty}
  c^{(a)}_k\sin(q_{x,2k}x)
\end{equation}
for antisymmetric modes, we arrive at the eigenvalue problem
similar to Eqs.~(\ref{Eq-EigenvalueProblem-GC-S}) and 
(\ref{Eq-EigenvalueProblem-GC-A}) [only a difference is that 
the indices $k$ and $k'$ start from $1$ in 
Eq.~(\ref{Eq-EigenvalueProblem-GC-S})]
with the matrix elements
\begin{eqnarray}\label{Eq-MatrixElement-SD-S}
 \theta^{(s)}_{kk'}
  & = & 
  \frac{2}{\pi}\left[(q_{2k'-1}l)^2+(q_yl)^2\right]
  \int_{-1}^{1}d\xi\int_{-1}^{1}d\xi'
  G^{(SD)}(\xi, \xi')
  \nonumber \\
 & & \times \cos(q_{2k-1}l\xi)\cos(q_{2k'-1}l\xi')
\end{eqnarray}
for symmetric modes and
\begin{eqnarray}\label{Eq-MatrixElement-SD-A}
 \theta^{(a)}_{kk'}
  & = &
  \frac{2}{\pi}\left[(q_{x,2k'}l)^2+(q_yl)^2\right]
  \int_{-1}^{1}d\xi\int_{-1}^{1}d\xi'
  G^{(SD)}(\xi, \xi')
  \nonumber \\
 & & \times \sin(q_{x,2k}l\xi)\sin(q_{x,2k'}l\xi').
\end{eqnarray}

\begin{figure}[ht]
 \begin{center}
  \includegraphics[height=6cm]{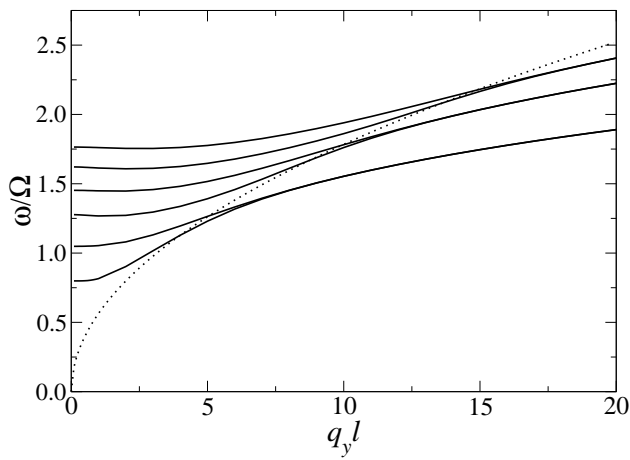}
  \caption{\label{Fig-dispersionSD} Dispersion relation for
  a slot diode with $q_n l/\pi = 1/2,\, 1,\, 3/2,\, 2,\,
  5/2,\, 3$. 
  Solid lines with lowest frequencies through highest ones
  correspond to $q_nl/\pi = 1/2$ through $q_nl/\pi = 3$.
  The dotted line is the dispersion relation of infinite 2D plasmons.}
 \end{center}
\end{figure}

\begin{figure}[ht]
 \begin{center}
  \includegraphics[height=6.8cm]{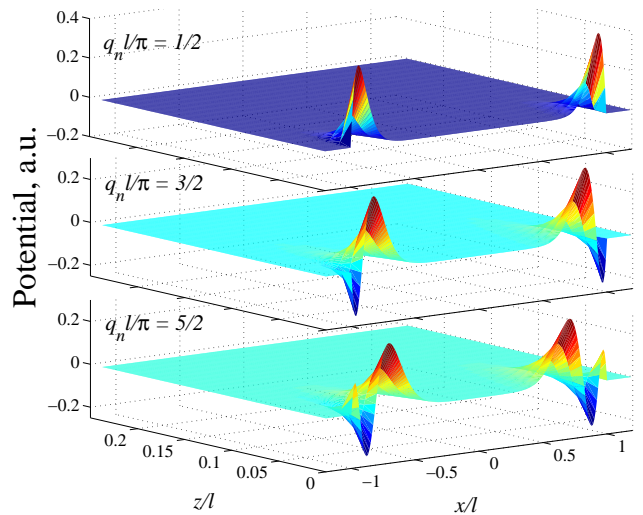}
  \caption{\label{Fig-potSD} Potential distributions of 
  three symmetric modes ($q_nl/\pi = 1/2, 3/2,$ and $5/2$.) 
  with $q_yl = 30$. These strip modes have two, four, and six peaks, 
  respectively.}
 \end{center}
\end{figure}

Solving the eigenvalue problems numerically and 
substituting eigenvalues into Eq.~(\ref{Eq-DispersionRelation}),
dispersion relation for a slot diode is obtained 
(Fig.~\ref{Fig-dispersionSD}). Frequencies in the limit
$q_yl\to0$ coincide with those calculated for standing
plasma waves in a slot diode in Ref.~\onlinecite{Satou-JJAP-2005}. 
Aside from the first modes, we have found that
higher strip modes can exist. By observing the potential
distributions, it was turned out that the potential
distribution of a mode begins to change and be localized near 
the edges of the 2DEG strip at the wave number $q_y$ at which
the dispersion relation curve of the mode crosses over
the curve of infinite 2D plasmons. One can also see from 
Fig.~\ref{Fig-dispersionSD} that
dispersion relation curves of symmetric and antisymmetric modes
with same index begin to coincide at this wave number.
Figure~\ref{Fig-potSD} shows potential distributions of
symmetric modes for $q_nl/\pi = 1/2, 3/2$, and $5/2$
with $q_yl = 30$. From Fig.~\ref{Fig-potSD}, one can see 
the increasing of peak numbers as $q_nl/\pi$ increases,
and peaks become larger and wider as they are farther from
an edge of the 2DEG strip.

\section{Conclusions}

In summary, we have studied plasmons in the following
2D systems:
(a) the 2DEG strip with a highly conducting gate (gated cavity)
and (b) the 2DEG strip with semi-infinite highly conducting
side contacts (slot diode).
We have found the dispersion relations of the plasmons and
the spatial distributions of ac electric potential.
In particular,
it has been shown that at sufficiently large wave numbers of
plasmons, their ac electric potential of some modes
can be localized near both edges of the strip, and their 
frequencies are lower than those of infinite 2D plasmons.

\section*{Acknowledgements}

The authors are thankful to A. Chaplik and M. S. Shur for
numerous discussions.

\appendix*
\section{}

Eqs.~(\ref{Eq-MatrixElement-GC-1}) through
(\ref{Eq-MatrixElement-GC-3}) can be reduced to the following
simpler expressions which contain only single integrals:
\begin{eqnarray}\label{Eq-MatrixElement-GC-6}
 \theta^{(s)}_{00} & = & \frac{2}{\pi}(q_yl)^2
  \left\{
   K_0(2q_yl)\mathcal{L}_{-1}(2q_yl)+K_1(2q_yl)\mathcal{L}_{0}(2q_yl)
  \right.
  \nonumber \\
 & &
  \left.
   -\left[ 1-2q_ylK_1(2q_yl) \right]/2\pi(q_yl)^2
   \right.
  \nonumber \\
 & &
  \left.
  -\frac{1}{\pi}\int_{0}^{2}d\xi
  \left(1-\frac{\xi}{2}\right)K_0\left(q_yl\sqrt{\xi^2+a^2}\right)
  \right\},
\end{eqnarray}
where $K_n$ is the modified Bessel function of second kind,
$\mathcal{L}_n$ is the modified Struve function~\cite{MathFunction}, 
and $a = 2W/l$,
\begin{equation}\label{Eq-MatrixElement-GC-7}
 \theta^{(s)}_{k0} = (-1)^{k+1}\frac{2\sqrt{2}}{\pi^2 k}(q_yl)^2
  \int_{0}^{2}G^{(GC)}(\xi)\sin(k\pi\xi),
\end{equation}
and
\begin{eqnarray}\label{Eq-MatrixElement-GC-8}
 \theta^{(s)}_{0k'} & = & (-1)^{k'+1}\frac{2\sqrt{2}}{\pi^2 k'}
  [(k'\pi)^2+(q_yl)^2]
  \nonumber \\
 & & \times \int_{0}^{2}G^{(GC)}(\xi)\sin(k'\pi\xi).
\end{eqnarray}
Here, $G^{(GC)}(\xi) = G^{(GC)}(\xi, 0)$.
Eq.~(\ref{Eq-MatrixElement-GC-4}) can be reduced to
\begin{equation}\label{Eq-MatrixElement-GC-9}
 \theta^{(s)}_{kk} = -\frac{2[(k\pi)^2+(q_yl)^2]}{k\pi^2}
  \int_{0}^{2}d\xi S_{k}(\xi)
\end{equation}
and
\begin{equation}\label{Eq-MatrixElement-GC-10}
 \theta^{(s)}_{kk'} = (-1)^{k+k'+1}
  \frac{4[(k'\pi)^2+(q_yl)^2]}{(k^2-k'^2)\pi^3}[T_{k}(2)-T_{k'}(2)]
\end{equation}
when $k \neq k'$. Similarly, 
Eq.~(\ref{Eq-MatrixElement-GC-5}) can be reduced to
\begin{equation}\label{Eq-MatrixElement-GC-11}
 \theta^{(a)}_{kk} = -\frac{2[(h\pi)^2+(q_yl)^2]}{h\pi^2}
  \int_{0}^{2}d\xi S_{h}(\xi)
\end{equation}
and
\begin{equation}\label{Eq-MatrixElement-GC-12}
 \theta^{(a)}_{kk'} = (-1)^{k+k'+1}
  \frac{4[(h'\pi)^2+(q_yl)^2]}{(h^2-h'^2)\pi^3}[T_{h}(2)-T_{h'}(2)],
\end{equation}
where $h = (2k-1)/2$ and $h' = (2k'-1)/2$, when $k \neq k'$.
Here,
\begin{equation}\label{Eq-MatrixElement-GC-13}
 S_{\nu}(\xi) = \int_{0}^{\xi}d\xi'\sin(\nu\pi\xi')
  \frac{d}{d\xi'}G^{(GC)}(\xi')
\end{equation}
and
\begin{equation}\label{Eq-MatrixElement-GC-14}
 T_{\nu}(\xi) = \int_{0}^{\xi}d\xi'[\cos(\nu\pi\xi')-1]
  \frac{d}{d\xi'}G^{(GC)}(\xi').
\end{equation}
Since all integrands of integrations in 
Eqs.~(\ref{Eq-MatrixElement-GC-6}) through
(\ref{Eq-MatrixElement-GC-14}) have no singularity, 
the Gaussian quadrature~\cite{NumericalAnalysis} was used to
perform the numerical integrations.

On the other hand, since Eq.~(\ref{Eq-GreenFunctionSD})
is very complicated, we directly performed double integrations
in Eqs.~(\ref{Eq-MatrixElement-SD-S}) and 
(\ref{Eq-MatrixElement-SD-A}) after integrating them by parts:
\begin{eqnarray}\label{Eq-MatrixElement-SD-S-2}
 & &
 \theta^{(s)}_{kk'}
   =  \frac{2q_{2k-1}l}{\pi}\left[(q_{2k'-1}l)^2+(q_yl)^2\right]
  \nonumber \\
 & & 
 \times
  \int_{-1}^{1}d\xi\int_{-1}^{1}d\xi'
  \mathcal{G}^{(SD)}(\xi, \xi')\sin(q_{2k-1}l\xi)\cos(q_{2k'-1}l\xi')
  \nonumber \\
\end{eqnarray}
and
\begin{eqnarray}\label{Eq-MatrixElement-SD-A-2}
 & & \theta^{(a)}_{kk'}
  = 
  -\frac{2q_{2k}l}{\pi}\left[(q_{2k'}l)^2+(q_yl)^2\right]
  \nonumber \\
 & & \times
  \int_{-1}^{1}d\xi\int_{-1}^{1}d\xi'
  \mathcal{G}^{(SD)}(\xi, \xi')\cos(q_{2k}l\xi)\sin(q_{2k'}l\xi'),
  \nonumber \\
\end{eqnarray}
where 
$\mathcal{G}^{(SD)}(\xi, \xi') = \int_{-1}^{\xi}G^{(SD)}(t, \xi')dt$. 
Since $G^{(SD)}$ has logarithmic singularity at $\xi = \xi'$, 
$\mathcal{G}^{(SD)}$ is the continuous function of $\xi$ and $\xi'$.
$\mathcal{G}^{(SD)}$ was calculated using 
adaptive quadrature~\cite{NumericalAnalysis}.
Finally, Eqs.~(\ref{Eq-MatrixElement-SD-S-2}) and
(\ref{Eq-MatrixElement-SD-A-2}) were calculated by 
Simpson's double integral method~\cite{NumericalAnalysis}.

\end{document}